\documentclass[prx,twocolumn,superscriptaddress,showpacs,amsmath,amstex,amssymb,citeautoscript,10pt]{revtex4-2}
\pdfoutput=1 

\usepackage{dcolumn}
\usepackage{bm}
\usepackage[T1]{fontenc}
\usepackage{braket}
\usepackage{graphicx}
\usepackage{xcolor}
\usepackage{lipsum}
\usepackage[normalem]{ulem}
\usepackage{amsmath}
\usepackage{amsfonts}
\usepackage{amssymb} 
\usepackage{color}
\usepackage[percent]{overpic}
\usepackage{soul} 
\usepackage{amssymb}
\usepackage{wasysym}
\usepackage{dsfont}
\usepackage{float}
\usepackage[mathscr]{euscript}
\usepackage{ifthen}
\usepackage{csquotes}
\usepackage{appendix}
\usepackage[bottom]{footmisc}
\usepackage{verbatim}
\usepackage{multirow}
\usepackage{mathrsfs}
\usepackage[scr=esstix,cal=boondox]{mathalfa} 
\usepackage{CJKutf8}
\usepackage{lineno}
\renewcommand{\vec}[1]{\mathbf{{#1}}}

\begin{document}

\title{
Observation of spin squeezing with contact interactions in one- and three-dimensional easy-plane magnets}

\author{Yoo Kyung Lee}
\altaffiliation{These authors contributed equally to this work.}
\affiliation{Department of Physics, Massachusetts Institute of Technology, Cambridge, MA 02139, USA}
\affiliation{Research Laboratory of Electronics, Massachusetts Institute of Technology, Cambridge, MA 02139, USA}
\affiliation{MIT-Harvard Center for Ultracold Atoms, Cambridge, MA, USA}

\author{Maxwell Block}
\altaffiliation{These authors contributed equally to this work.}
\affiliation{MIT-Harvard Center for Ultracold Atoms, Cambridge, MA, USA}
\affiliation{Department of Physics, Harvard University, Cambridge, MA 02138, USA}

\author{Hanzhen Lin
(\begin{CJK*}{UTF8}{gbsn}林翰桢\end{CJK*})}
 \altaffiliation{These authors contributed equally to this work.}
\affiliation{Department of Physics, Massachusetts Institute of Technology, Cambridge, MA 02139, USA}
\affiliation{Research Laboratory of Electronics, Massachusetts Institute of Technology, Cambridge, MA 02139, USA}
\affiliation{MIT-Harvard Center for Ultracold Atoms, Cambridge, MA, USA}

\author{{Vitaly Fedoseev}}
\altaffiliation{These authors contributed equally to this work.}
\affiliation{Department of Physics, Massachusetts Institute of Technology, Cambridge, MA 02139, USA}
\affiliation{Research Laboratory of Electronics, Massachusetts Institute of Technology, Cambridge, MA 02139, USA}
\affiliation{MIT-Harvard Center for Ultracold Atoms, Cambridge, MA, USA}

\author{\\Philip J. D. Crowley}
\affiliation{MIT-Harvard Center for Ultracold Atoms, Cambridge, MA, USA}
\affiliation{Department of Physics, Harvard University, Cambridge, MA 02138, USA}

\author{Norman Y. Yao}
\affiliation{MIT-Harvard Center for Ultracold Atoms, Cambridge, MA, USA}
\affiliation{Department of Physics, Harvard University, Cambridge, MA 02138, USA}

\author{Wolfgang Ketterle}
\affiliation{Department of Physics, Massachusetts Institute of Technology, Cambridge, MA 02139, USA}
\affiliation{Research Laboratory of Electronics, Massachusetts Institute of Technology, Cambridge, MA 02139, USA}
\affiliation{MIT-Harvard Center for Ultracold Atoms, Cambridge, MA, USA}

\begin{abstract}
\noindent
\noindent 

Entanglement in a many-particle system can enable measurement sensitivities beyond that achievable by only classical correlations. 
For an ensemble of spins, all-to-all interactions are known to reshape the quantum projection noise, leading to a form of entanglement known as spin squeezing. 
Here, we demonstrate spin squeezing with strictly short-range contact interactions. 
In particular, working with ultracold lithium atoms in optical lattices, we utilize superexchange interactions to realize a nearest-neighbor anisotropic Heisenberg model.
We investigate the resulting quench dynamics from an initial product state in both one and three dimensions. 
In 1D, we observe $1.9^{+0.7}_{-0.5}$ dB of spin squeezing in quantitative agreement with theory. 
However, in 3D, we observe a maximum of $2.0^{+0.7}_{-0.7}$ dB of squeezing, over an order of magnitude smaller than that expected. 
We demonstrate that this discrepancy arises from the presence of a finite density of holes;  both the motion of the holes as well as direct coupling between spin and density qualitatively alter the spin dynamics. 
Our observations point to the importance of understanding the complex interplay between motional and spin degrees of freedom in quantum simulators. 

\end{abstract}

\maketitle
Entanglement is a unique feature of correlated quantum systems, which can enable enhanced capabilities in settings ranging from computing to metrology \cite{expt_ions_2001,expt_ions_ghz_2004,ligo_2013,rydberg_array_2022,entanglement_2024}.
However, preparing large-scale entanglement in many particle systems remains a challenging frontier. 
Indeed, an environmental perturbation on even a single particle can, in principle, collapse an entire wave function onto a mixed state, exhibiting only classical correlations. 

In strongly interacting spin ensembles, a paradigmatic example of  metrologically-useful entanglement is so-called spin-squeezing, where the variance of a global spin operator is smaller than allowed by the standard quantum limit (SQL). 
While the prototypical setting for observing such squeezing is in  all-to-all interacting systems~\cite{expt_cavity_2010,expt_cavity_2010_2,expt_cavity_2016,expt_cavity_2019,expt_cavity_2022,expt_cavity_2024}, spin squeezed states have also been realized with power-law interactions~\cite{expt_ions_2001,expt_rydberg_2023,expt_rydberg_browaeys_2023,expt_clock_kaufman_2023}.
On the other hand, many atomic platforms with metrological significance (e.g.~optical lattice clocks) exhibit only contact interactions \cite{revmodphys_clock_2011,hinkley_clock_2013,clock_2017,clock_2022}. 
Motivated by recent theoretical advances \cite{theory_shortrange_2020,maxbingtian}, we investigate spin squeezing in a nearest-neighbor anisotropic Heisenberg (XXZ) model, realized via a two-component Mott insulator near unity filling. 
Working in the easy-plane regime where spin-exchange interactions (i.e.~XY interactions) dominate, we observe a maximum of $1.9^{+0.7}_{-0.5}$ dB of spin squeezing in one-dimensional chains, in agreement with theory.

In three dimensions, nearest-neighbor XY interactions have recently been predicted to generate \emph{scalable} spin squeezing, where the metrological gain improves with system size owing to the presence of finite-temperature continuous symmetry breaking~\cite{maxbingtian,theory_roscilde_2024}. 
However, achieving such long-range order requires the temperature of an initial state to be lower than the critical ordering temperature, 
$T_\textrm{c} \sim J$.
While we observe spin squeezing of up to 2.0 dB in 3D, we do not recover the 16 dB expected from the scalable squeezing of $\sim10^4$ spins. 
Combined experimental and theoretical investigation suggests that the deviation arises due to the presence of holes (estimated to be $\sim10\%$), which increases the effective temperature of our initial state above the critical ordering temperature $T_\textrm{c}$. 
Moreover, previously unexplored direct coupling between the motional and spin degrees of freedom causes rapid decay of the spin length, which limits the maximum interaction times and thus, the amount of achievable spin squeezing.
We emphasize that this effect is present for all spin models realized via superexchange interactions, regardless of the atomic species. 
Thus, accounting for the presence of holes in such systems will be crucial for utilizing them as tunable quantum simulators. 

\begin{figure}[t]
\vspace{5pt}
    \centering
    \includegraphics[width=\linewidth,keepaspectratio]{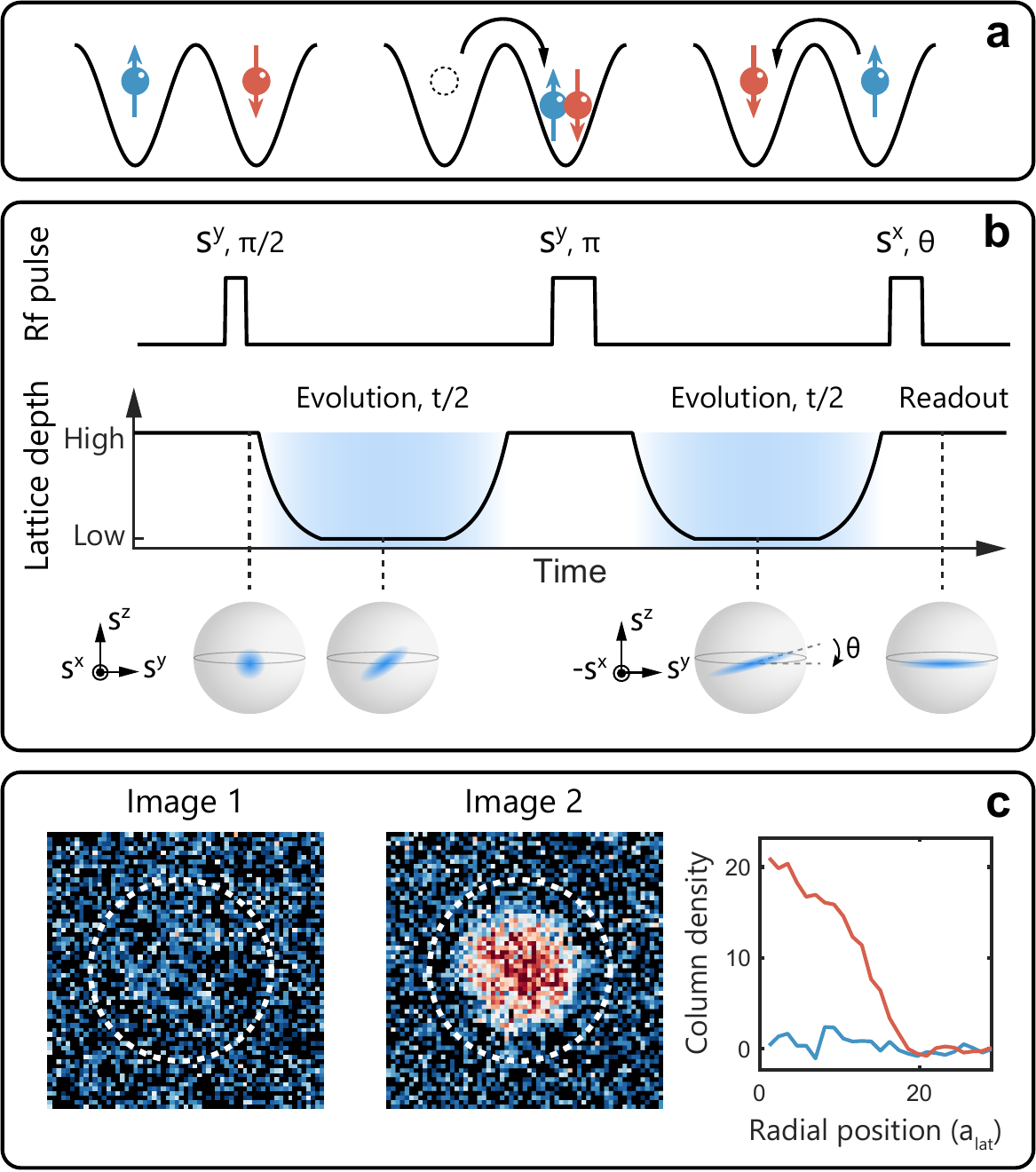}
    \caption{Experimental setup. (a) Two hyperfine states of $^7$Li atoms in the Mott insulating regime encode a spin-$1/2$ Heisenberg Hamiltonian [eq.~(\ref{eq:heis})]. In this mapping, spin-spin interactions are moderated by superexchange, a second-order tunneling process \cite{duan_spinexchange_2003}. (b) After a global $\pi/2$ pulse to rotate the spins into the easy-plane, we reduce the lattice depth in one or three directions and allow the spins to interact. Under Heisenberg interactions, the spin distribution shears and the variance of the initial state is redistributed along different axes. A spin echo is included halfway through the evolution time to mitigate effects from magnetic field fluctuations. We raise the lattice depth, then rotate the spins along the mean spin direction by angle $\theta$ before measurement. (c) Left: After the rotation, polarization contrast imaging \cite{jepsen_nature_2020} is used to directly measure the global spin imbalance; this probes the spin operator $S^\theta$. Center: The same sample is exposed to a second imaging pulse where the light is detuned with respect to the first image, confirming that there are atoms present. The second image is used to fit the position of the cloud. Right: Radial averages of the profiles in image 1 (blue line) and image 2 (red line).
    }
    \label{fig:experiment}
\end{figure}

\emph{Experiment}.--- 
We realize spin models by loading two hyperfine states of ultracold $^7$Li atoms into a cubic optical lattice \cite{duan_spinexchange_2003,jepsen_nature_2020,jepsen_prx_2021,jepsen_phantom_2022}.
This system is well-characterized by a Bose-Hubbard model. 
In the Mott insulating regime at unity filling, it maps onto a spin-1/2 XXZ model
\begin{equation}
    H_0 = \sum_{\langle ij \rangle} [J (S^x_iS^x_j + S^y_iS^y_j) + J_z S^z_iS^z_j] \label{eq:heis}
\end{equation}
where the indices $\langle ij\rangle$ run over nearest neighbors, the spin degree of freedom $\ket{\uparrow},\ket{\downarrow}$ is encoded by the hyperfine states, and $S^{x,y,z}_i$ are the single-particle spin operators. 
Both the spin-exchange $J=-4\tilde{t}^2/U_{\uparrow\downarrow}$ and Ising $J_z=4\tilde{t}^2/U_{\uparrow\downarrow}-4\tilde{t}^2/U_{\uparrow\uparrow}-4\tilde{t}^2/U_{\downarrow\downarrow}$ couplings are mediated by superexchange (Fig.~\ref{fig:experiment}a) and can be tuned using Feshbach resonances \cite{amatogrill_spectroscopy_2019}.
Here, $\tilde{t}$ is the tunneling matrix element and $U_{\sigma \sigma'}$ are the on-site interaction energies and the subscript $\sigma=\uparrow,\downarrow$ refers to the spin degree of freedom \cite{supp}.
In the easy-plane XY regime (with $|J| > |J_z|$), the dynamics of $H_0$ shear the uncertainty (projection noise) of an initial spin-polarized state, leading to squeezing (Fig.~\ref{fig:experiment}b)~\cite{maxbingtian}.

The experimental sequence used to generate spin squeezing is depicted in Fig.~\ref{fig:experiment}b. We initialize the atoms in a deep optical lattice at $37E_\textrm{R}$ where spin dynamics are negligible. Here, $E_\textrm{R}/\hbar=\hbar k^2/2m\approx 2\pi\times25\,{\rm kHz}$ is the recoil energy for an atom of mass $m$ subject to a lattice formed with laser beams of wave-vector $k=\pi/a_{\rm lat}$. The lattice spacing is $a_{\rm lat}=532$ nm and $\hbar$ is the reduced Planck constant.
Beginning with a spin-polarized initial state along $S^z$ $\ket{\uparrow}^{\otimes N}$, we utilize  a global $\pi/2$ pulse around the $S^y$ axis to prepare the $S^x$-polarized state $[(\ket{\uparrow} + \ket{\downarrow})/\sqrt{2}]^{\otimes N}$. 
Next, we turn on the Hamiltonian, $H_0$, by reducing the lattice depth. 
Using the known scattering lengths \cite{amatogrill_spectroscopy_2019,eindhoven_scattering_2011}, we infer that our system realizes an anisotropy in the XY regime with $J_z/J=-0.18$~\cite{footnote_ferromagnetic}.
After a variable evolution time $t$ which includes a spin-echo pulse, we raise the lattice to freeze the dynamics, and then rotate the system about its mean spin direction by a variable angle $\theta$. 

The improvement in metrological sensitivity relative to the initial product state is characterized by the squeezing parameter~\cite{wineland_1992,wineland_1994},
\begin{equation}
    \xi^2 \equiv \frac{N {\rm min}_\theta\left({\rm Var}\,\left[S^\theta\right]\right)}{\langle S^x \rangle^2} 
    \label{eq:squeezing}
\end{equation}
where $\langle S^x \rangle$ is the mean spin length and ${\rm min}_\theta$ indicates the minimum with respect to $\theta$. 
Thus, by measuring both $\langle S^x\rangle$ and the variance of $S^{\theta}=\sum_{i=1}^N \left[\cos(\theta)S_i^z + \sin(\theta)S_i^y\right]$ (for multiple angles $\theta$) as a function of time $t$, one can directly probe the dynamics of the squeezing parameter. 
We note that the SQL corresponds to $\xi^2=1$ and that $\xi^2<1$ is a witness for entanglement \cite{entanglement_proof_2001}. 

As depicted in Fig.~\ref{fig:experiment}c, polarization contrast imaging \cite{jepsen_nature_2020} is used to measure the spin imbalance, $S^z=\frac{1}{2}(N_\uparrow-N_\downarrow)$, after the variable rotation, thereby providing an effective measurement of $S^\theta$ \cite{supp}. The states we use are magnetic field-sensitive, so we include a spin-echo pulse in the middle of the evolution to mitigate the effects of slowly-varying magnetic field fluctuations (Fig.~\ref{fig:experiment}b). 
Furthermore, we divide the cloud into two subsystems, $a$ and $b$, with $N_a,  N_b \approx 10^4$, and then analyze the difference in $S^z$ between them while subtracting the photon shot noise \cite{supp}. The variance of this difference can be used to approximate the total variance with negligible technical noise, i.e.~${\rm Var}[S^\theta] \approx {\rm Var} [S^\theta_a-S^\theta_b]$. 
Henceforth, ${\rm Var}[S^\theta]$ refers to the variance measured using this subsystem analysis.
Without this method, we cannot perform measurements at the quantum limit, likely due to fluctuations in the precession angle between Ramsey pulses caused by magnetic field noise that is faster than the echo sequence. 

\emph{Results of squeezing in 1D and 3D}.---
\begin{figure}[t]
    \centering
    \vspace{-10pt}
    \includegraphics[width=\linewidth,keepaspectratio]{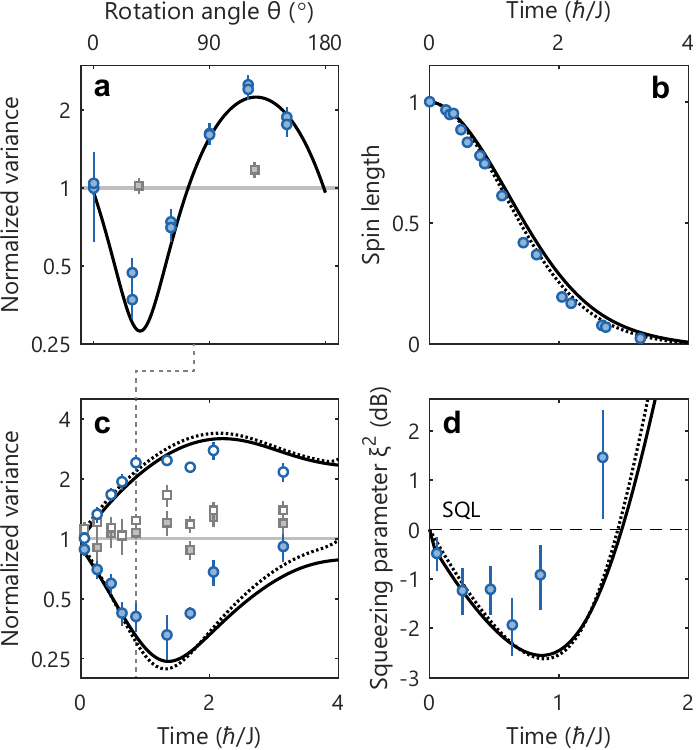}
    \caption{Spin squeezing in 1D chains. (a) The normalized variance $4 {\rm Var} [S^\theta] / N$ is measured as a function of final rotation angle $\theta$. Shown is an example at time $t=0.86\,\hbar/J$. Gray squares indicate results where atoms are held in a deep lattice with negligible superexchange interactions and the horizontal gray line is the theory for non-interacting atoms. (b) The spin length $2\,\langle S^x\rangle/N$ decays to zero within several $\hbar/J$ due to the absence of ordering. (c) The minimum (closed) and maximum (open symbols) variances vs. $\theta$ for interacting (circles) and non-interacting (squares) samples. (d) Using eq.~(\ref{eq:squeezing}), we determine the squeezing parameter. The region below the dashed line labeled SQL indicates values of $\xi^2$ which signify entanglement. We measure a maximum of 1.9 dB of squeezing \cite{supp}. All error bars represent one standard deviation resulting from jackknife estimation. The simulations are performed with 32 spins. In all panels, the solid (dotted) line is a theoretical curve with a 5\% (0\%) hole density, with all other parameters fixed by experiment \cite{supp}.
    }
    \label{fig:results1D}
\end{figure}
After preparation of the $S^x$-polarized state, we turn on the spin dynamics by reducing the lattice depth.
To realize 1D chains, the lattice is reduced in only one direction to $13\,E_\textrm{R}$, which yields dynamics under $H_0$ with a spin-exchange rate of $J/\hbar=2\pi\times38$ Hz. 
Our results are summarized in Fig.~\ref{fig:results1D}.
For each interaction time $t$, we measure both the normalized variance $4 {\rm Var}\,[S^\theta]/N$ and the spin length $2\langle S^x \rangle/N$: for instance, Fig.~\ref{fig:results1D}a depicts the variance at time $t = 0.86\,\hbar/J$, and Fig.~\ref{fig:results1D}b the spin length for times up to $4\hbar/J$. We choose this normalization so that a polarized product state has a normalized variance and spin length of 1. The minimum (squeezing) and maximum (anti-squeezing) values of the variance as a function of time are shown in Fig.~\ref{fig:results1D}c, as the closed (open) circles. 
By contrast, for a non-interacting sample where the lattice is kept high (closed and open gray squares, Fig.~\ref{fig:results1D}c),  the minimum and maximum variance always remains near unity. 

The dotted black lines are the results of TDVP (time dependent variational principle \cite{ITensor-r0.3}) 
simulations (Fig.~\ref{fig:results1D}), while the solid lines include the effect of a small density of holes, which is described in more detail below [see eqs.~(\ref{eq:tunneling}) and (\ref{eq:tj_correction})].
The rapid decay of the mean spin length to zero, as observed in Fig.~\ref{fig:results1D}b, is consistent with the theoretical expectation that  there is no long-range XY order for nearest-neighbor interactions in 1D \cite{supp,Sachdev_2011}.
Nevertheless, for early times $t<2\,\hbar/J$, we find that the variance decreases faster than the mean spin length, leading to spin squeezing (Fig.~\ref{fig:results1D}d); $\xi^2$ is optimized at time $t \approx 0.64 \hbar/J$ where we realize $\approx 1.9$ dB of spin squeezing, in close agreement with theory. 

\begin{figure}[t]
    \centering
    \includegraphics[width=\linewidth,keepaspectratio]{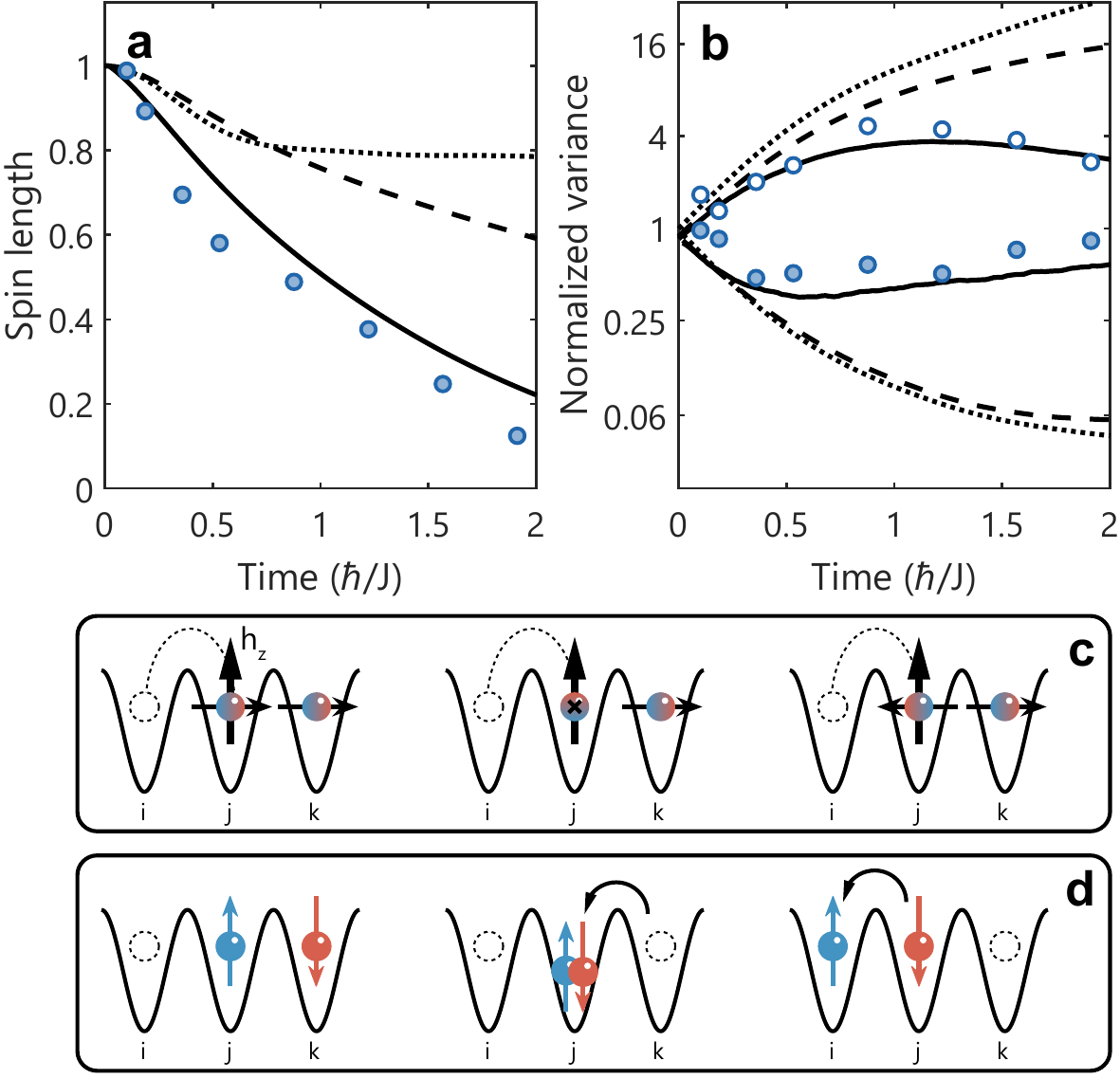}
    \caption{Spin squeezing in 3D. Decay of spin length (a) and the minimum and maximum variances (b) as a function of time in 3D. Also shown are simulations for 10,648 spins with no holes (dotted curve), with $\sim 10\%$ hole fraction evolving under a Hamiltonian with just $H_{\rm t}$ (dashed), and the addition of both $H_{\rm t}$ and $H_{\rm d}$ (solid). The gap between theory and experiment can only be explained by the inclusion of the terms in eq.~(\ref{eq:tj_correction}). (c)-(d) Illustration of explicit spin-density coupling terms $H_{\rm d}$ in eq.~(\ref{eq:tj_correction}). (c) The effective magnetic field $\propto h_z$ dephases atoms with neighboring holes. Two spins which are initially aligned (left) will dephase (center, right) due to the presence of the hole next to a spin. (d) The spin-flip-assisted tunneling term is a virtual process where the spin which tunnels two sites flips its spin, along with the spin of its neighbor. }
    \label{fig:results3D}
\end{figure}

To explore the dynamics of spin squeezing in 3D, we repeat the same protocol but instead ramp all three lattices from $37\,E_\textrm{R}$ to $15\,E_\textrm{R}$, resulting in a spin-exchange coupling of $J/\hbar=2\pi\times27$ Hz.
Fig.~\ref{fig:results3D} compares our experimental measurements of the contrast decay and variance as a function of time, with numerical simulations utilizing the discrete truncated Wigner approximation (DTWA)~\cite{maxbingtian,dtwa_rey_2015}. 
Unlike in 1D, for $\langle S^x\rangle$, the theory (dotted black line) exhibits a small transient decay, followed by a plateau for $t \gtrsim 0.7 \hbar /J$.
This plateau is indicative of the finite temperature long-range order that occurs for nearest-neighbor XY interactions in 3D. 
However, this expected behavior is not borne out in experiment. 
Rather, the spin length quickly decays without any sign of a plateau (Fig.~\ref{fig:results3D}a). 
This discrepancy between theory and experiment also extends to the dynamics of the variance (Fig.~\ref{fig:results3D}b).
In particular, for a system of $\sim 10^4$ spins,
one expects $4 {\rm Var}[S^\theta]/N$ to quickly reduce to $0.04$ at time $t=2\hbar/J$ (theory, dotted black line); however, the minimum variance we observe is significantly larger, $0.47$. 

We conjecture that the above discrepancies arise from a subtle, but crucial, coupling between the spin and density degrees of freedom.
In particular, owing to the finite temperature of our BEC and imperfect lattice loading, we expect a $\sim 10\%$ hole fraction in our system \cite{supp}; this density entropy can then couple to the spin sector via higher-order terms in the effective  Hamiltonian, thereby destabilizing long-range XY order. 

\emph{Spin-density coupling due to holes}.---In the presence of holes, the Hamiltonian is modified to $H=H_0+H_\textrm{t}+H_\textrm{d}$ \cite{jepsen_prx_2021}, where
\begin{align}
    H_\textrm{t}&=-\tilde{t}\sum_{\langle ij\rangle,\sigma}a_{i\sigma}^\dagger a_{j\sigma} + {\rm H.c.} \label{eq:tunneling}\\
    H_\textrm{d} &= \sum_{\langle ij \rangle} \frac{h_z}{2}\left[S_i^z(n_{j,\uparrow} + n_{j,\downarrow})+(n_{i,\uparrow}+n_{i,\downarrow})S_j^z\right]\nonumber\\
    &+\sum_{\langle ijk\rangle,\sigma}\frac{\tilde{t}^2}{U_{\uparrow\downarrow}}a_{i \bar{\sigma}}^\dagger S_j^\sigma a_{k\sigma} + \rm{H.c.}  
    \label{eq:tj_correction}
\end{align}
Here, $a_{i\sigma}^\dagger (a_{i\sigma})$ corresponds to the atomic creation (annihilation) operator on site $i$, $\sigma$ and $\bar{\sigma}$ represent opposite spin states, and $\langle ijk \rangle$ represent three neighboring sites. $h_z\,{=}\,4\tilde{t}^2(1/U_{\uparrow\uparrow}-1/U_{\downarrow\downarrow}) = -1.1J$ according to measured scattering lengths, and $S_i^{\uparrow (\downarrow)} \equiv S_i^{+(-)}$ are the spin raising (lowering) operator on each site.
$H_{\rm t}$ describes the hopping of holes, while $H_{\rm d}$ captures the explicit coupling between spin and density. 
The first term of $H_{\rm d}$ corresponds to an effective magnetic field, whose strength  depends on the density of neighboring lattice sites; we note that the coefficient of this term, $h_z$, can in principle, be tuned to zero via Feshbach resonances and the choice of atomic species~\cite{duan_spinexchange_2003,jepsen_nature_2020,jepsen_prx_2021}.
Meanwhile, the second term of $H_{\rm d}$ captures spin-flip-assisted tunneling (Fig.~\ref{fig:results3D}d) \cite{jepsen_prx_2021}. 
We note that all terms in $H_{\rm d}$ are on the same order as $J\sim\tilde{t}^2/U_{\sigma\sigma'}$, but  are suppressed by the hole probability and  do not contribute to spin dynamics in the absence of holes.
However, since the energy scale for localized density fluctuations ($\tilde{t}/\hbar\,=\,2\pi\times160$ Hz) greatly exceeds that of the spin-exchange $J$, even a small coupling between spin and density can decohere the spins and increase the effective spin temperature, to the detriment of long-range XY order and consequently, spin squeezing. 
To this end, we conjecture that the presence of these mobile holes causes the discrepancy between theory and experiment in Fig.~\ref{fig:results3D} and prevents us from seeing stronger amounts of spin squeezing in 3D.

Interestingly, this conjecture also explains the agreement between theory and experiment in 1D (Fig.~\ref{fig:results1D}).
In particular, as aforementioned, quantum fluctuations in 1D are too strong for long-range correlations to form, and there is no finite critical temperature for XY order. 
Consequently, even an ideal 1D system will \emph{not} exhibit a plateau in the contrast, nor scalable spin squeezing at fixed anisotropy $J_z/J$~\cite{theory_mott_2024}. 
Thus, any additional decoherence due to holes is overshadowed by this intrinsic decay, and plays a minor role. 
Furthermore, the spin-density terms in $H_{\rm d}$ couple a spin to its nearest and next-nearest neighbors.
Compared to 3D, a 1D system exhibits lower coordination (i.e.~three times fewer neighbors and eight times fewer next-nearest neighbors), which leads to less decoherence.

\emph{Theoretical and experimental exploration of holes}.---
\begin{figure}[b]
    \centering
    \includegraphics[width=\linewidth,keepaspectratio]{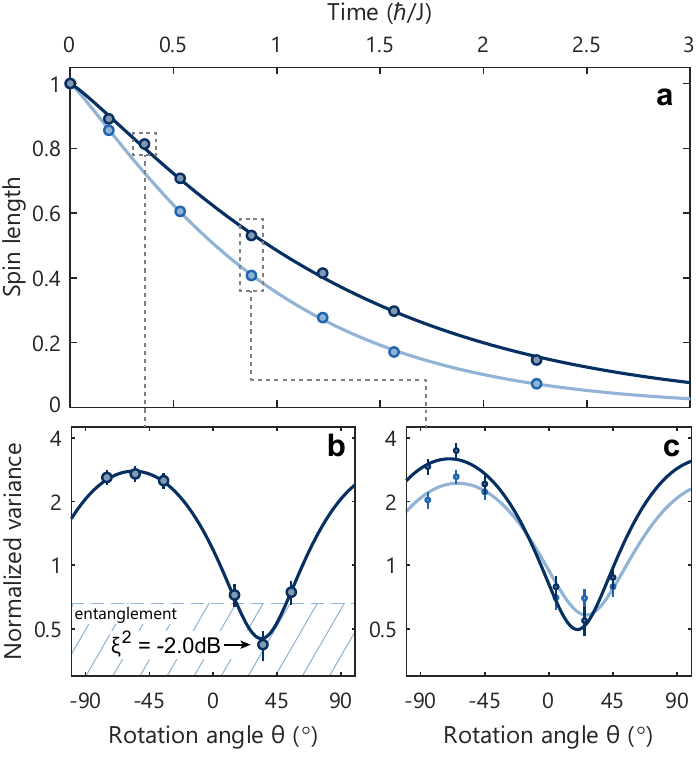}
    \caption{Squeezing improvement with colder 3D samples.
    (a) The spin length of colder samples (dark blue) decays slower than that of the normal sample (light blue). The solid lines are fits to guide the eye. (b) The variance as a function of rotation angle $\theta$ at time $t\,=\,0.36\,\hbar/J$. The hatched region denotes where the normalized variance is less than the squared spin length, demonstrating entanglement. With a cold sample, we achieve spin squeezing of 2.0 dB. (c) Comparing normalized variance of colder (dark blue) and normal (light blue) samples at time $t\,=\,0.88\,\hbar/J$. Both squeezing and anti-squeezing are enhanced for the colder sample.
    }
    \label{fig:temperature}
\end{figure}
To go beyond a qualitative analysis of the spin-density coupling, we utilize DTWA to numerically simulate the squeezing dynamics in the presence of holes.
To begin, we include the effect of $H_{\rm t}$ by augmenting DTWA with ``holes'' -- vacant sites that hop randomly to nearest neighbors during the simulation.
Although $H_{\rm t}$ does not directly couple the spin and density degrees of freedom, the spins must ``reshuffle'' as the holes move throughout the sample. 
Even with a modest hole fraction of $\sim 10\%$, this implicit coupling already has a noticeable effect: as illustrated in Fig.~\ref{fig:results3D}a (dashed line), the spin length decays rather than equilibrating to a constant, reflecting the fact that the holes have heated the system above the ordering temperature.
In contrast, for our system size, the implicit coupling to holes has little effect on the variance (Fig.~\ref{fig:results3D}b, dotted and dashed lines) \cite{footnote_system_sizes}.
Thus, the inclusion of $H_{\rm t}$ alone is not sufficient to explain our observed limitations on spin squeezing.

To this end, we now investigate the effect of the explicit spin-density coupling described in $H_{\rm d}$.
In particular, we account for the first term in $H_{\rm d}$ by adding a local magnetic field on sites adjacent to a hole (Fig.~\ref{fig:results3D}c); to account  for the second term, we apply a random rotation (in the $xy$-plane) to any spin that a hole hops over (Fig.~\ref{fig:results3D}d).
Heuristically, this latter operation is a semi-classical approximation of spin-flip-assisted tunneling and assumes that spin and density are disentangled. 

With the addition of both terms $H_{\rm t}$ and $H_{\rm d}$, the DTWA simulations exhibit remarkable agreement with the experimental results (Fig.~\ref{fig:results3D}, solid black curves). 
A few remarks are in order. 
First, the numerics correctly predict that squeezing is more adversely impacted by the presence of holes than anti-squeezing. 
Next, of the two terms in $H_{\rm d}$, the dominant effect on spin squeezing comes from the spin-flip-assisted tunneling~\cite{jepsen_prx_2021,supp}. 
Finally, our simulations (both TDVP and DTWA) confirm the intuition (and experimental data) that spin squeezing in 1D is not significantly affected by holes (Fig.~\ref{fig:results1D}).
This bolsters our hypothesis that previously neglected channels of spin-density coupling play a crucial role in the spin dynamics of 3D atomic ensembles.

To further test the role of spin-density coupling experimentally, we change the hole fraction with a different loading procedure (Fig.~\ref{fig:temperature}). 
In particular, we implement stronger evaporation in order to reach colder samples and reduce the experimental hole density. 
Using an identical spin squeezing protocol (Fig.~\ref{fig:experiment}), we observe a significantly  slower contrast decay, leading to quantitative enhancements in both the amount of squeezing and anti-squeezing. The strongest spin-squeezing we experimentally obtain with a colder sample is $2.0^{+0.7}_{-0.7}$ dB.

\emph{Conclusions and outlook}.---Our work opens the door to a number of intriguing directions.
First, it would be interesting to explore the use of entropy redistribution strategies~\cite{entropy_removal_2018,entropy_removal_2020} in order to reduce the  density of holes in a 3D Mott insulator, with the goal of observing scalable spin squeezing. 
Second, although our experiments have focused on 1D and 3D,  squeezing dynamics in 2D also exhibit a rich landscape to explore. Here, contact interactions lead to algebraic long-range order and a power-law decay of the spin length; theory suggests that for sufficiently low temperature initial states, correlations develop more quickly than the spin length decays, and scalable spin squeezing is possible \cite{theory_roscilde_2024}.
Third, our results point to the importance of quantifying the impact of holes on the spin dynamics.
Once characterized, one could further envision harnessing the interactions between spin and motion to realize an even richer array of Hamiltonians. 
Finally, our demonstration of spin squeezing with contact interactions, broadens the landscape of cold atomic experiments that can access and leverage such metrologically-useful entanglement. 

\emph{Acknowledgments}.---
We thank Servaas Kokkelmans for providing updated calculations of scattering lengths and Bingtian Ye for discussions. 
We acknowledge support from the NSF through grant No. PHY-2208004, from the Center for Ultracold Atoms (an NSF Physics Frontiers Center) through grant No. PHY-2317134, the Army Research Office (contract No. W911NF2410218 and MURI program grant no.~W911NF-20-1-0136), the US Department of Energy via BES grant no.~DE-SC0019241 and from the Defense Advanced Research Projects Agency (Grant No. W911NF2010090).
Y. K. L. is supported in part by the National Science Foundation Graduate Research Fellowship under Grant No. 1745302. 
Y. K. L and H. L acknowledge the MathWorks Science Fellowship.
N. Y. Y acknowledges support from a Simons Investigator award. 
{\bf Author contributions:}
Y. K. L., H. L., V. F., and W. K. conceived the experiment.
Y. K. L., H. L, and V. F. performed the experiment and analyzed the data. 
M. B. developed and performed the numerical simulations with input from P. C. and N. Y. Y.
All authors discussed the results and contributed to the writing of the manuscript.
{\bf Competing interests:} The authors declare no competing financial interests.
{\bf Data and materials availability:} The data that support the findings of this study are available from the corresponding author upon reasonable request.

\bibliographystyle{ieeetr}


\newpage
\clearpage

\setcounter{figure}{0}
\setcounter{equation}{0}
\makeatletter 
\renewcommand{\thefigure}{S\@arabic\c@figure}
\renewcommand{\theequation}{S\@arabic\c@equation}

\makeatother

\onecolumngrid

\section*{Supplemental material}

\section{Superexchange and Heinsenberg Interaction}
The Heisenberg interaction we realize comes from the superexchange processes of the native Bose-Hubbard Hamiltonian for atoms in an optical lattice. To understand such mapping colloquially, we can consider a two site model. We first consider an ideal Mott insulator with unity filling. In this case, the on-site interaction is much greater than tunneling; hence, the motion is mostly frozen out. The only allowed effect is of second order, where an atom tunnels to a neighboring filled site, experiences the onsite interaction, and tunnels back. This ``superexchange'' process contributes an energy shift on the order of $\tilde{t}^2/U$ to the Mott insulator state. Due to the three different scattering lengths available to $^7$Li in different  states ($a_{\uparrow\uparrow},a_{\uparrow\downarrow},a_{\downarrow\downarrow}$), the superexchange energy scales can be tuned with the magnetic field by the mapping \cite{duan_spinexchange_2003,jepsen_nature_2020}
\begin{align}
    J&=-4\tilde{t}^2/U_{\uparrow\downarrow} \\
    J_z&=4\tilde{t}^2(1/U_{\uparrow\downarrow}-1/U_{\uparrow\uparrow}-1/U_{\downarrow\downarrow}).
\end{align}
$J$ and $J_z$ are the transverse and Ising interactions in the Heisenberg Hamiltonian eq.~(\ref{eq:heis}), respectively. $U_{\sigma\sigma'}$ is the on-site energy of two spins and is given by the overlap of Wannier integral for two atoms on a site  $w(\vec{r})=w_x(x)w_y(y)w_z(z):$
\begin{align}
U_{\sigma\sigma'}&=\frac{4\pi\hbar^2a_{\sigma\sigma'}}{m}\int d^3\vec{r}|w(\vec{r})|^4.
\end{align}

However, the Mott insulators in the experiment have a finite hole fraction. In the presence of holes, the Bose-Hubbard maps onto a much richer Hamiltonian, which is derived in \cite{jepsen_prx_2021}:
\begin{equation}
\begin{split}
    H = & \sum_{\langle ij \rangle} \left\{J (S^x_iS^x_j + S^y_iS^y_j) + J_z S^z_iS^z_j -\frac{h_z}{2}[S^z_i(n_{j\uparrow}+n_{j\downarrow})+(n_{i\uparrow}+n_{i\downarrow})S^z_j] + c(n_{i\uparrow}+n_{i\downarrow})(n_{j\uparrow}+n_{j\downarrow})\right\}\\    
    & - \sum_{\langle ij \rangle, \sigma} \tilde{t}a^{\dagger}_{i\sigma}a_{j\sigma}+{\rm H.c.}- \sum_{\langle ijk \rangle, \sigma} \left( \frac{\tilde{t}^2}{U_{\uparrow\downarrow}} a^{\dagger}_{i\sigma}n_{j\bar{\sigma}}a_{k\sigma} + \frac{\tilde{t}^2}{U_{\uparrow\downarrow}} a^{\dagger}_{i\bar{\sigma}}S^{\sigma}_j a_{k\sigma} + \frac{2\tilde{t}^2}{U_{\sigma\sigma}} a^{\dagger}_{i\sigma}n_{j\sigma}a_{k\sigma}\right)+{\rm H.c}.
\end{split}
\label{eq:tj_full}
\end{equation}
The subscript $\sigma$ represents two spin states $\uparrow,\downarrow$ and $a_i^\dagger(a_i)$ are the creation (annihilation) operators of an atom on site $i$. The first two terms represent the desired Heisenberg interaction eq.~(\ref{eq:heis}) and the term proportional to $\tilde{t}$ the tunneling of holes. 

We emphasize that the presence of holes not only modifies the interaction structure through the missing sites or shuffles the spins, but also causes direct density-spin coupling in the following terms:

The term proportional to $h_z = 4\tilde{t}^2/U_{\uparrow\uparrow}-4\tilde{t}^2/U_{\downarrow\downarrow}$ represents the energy between $\ket{\uparrow}$ and $\ket{\downarrow}$ state, or an effective magnetic field that dependents on the number of nearest-neighbors a spin has. In an ideal Mott insulator of infinite size, each atom has the same number of neighbors, so this term only results in a global effective magnetic field. This is not the case, however, for spins that neighbor a hole or dwell at the boundary which have a different number of nearest neighbors; these spins are subject to dephasing with respect to the bulk. Generally speaking, the effective magnetic field terms may be cancelled or reduced by tuning the scattering lengths to the condition $a_{\uparrow\uparrow}=a_{\downarrow\downarrow}$. This was not accessible to our experiment given the states we used. This limited the available $|h_z|$ to $1-2J$; however, removing the $h_z$ term could be possible for other alkalis with different ratios of scattering lengths.

The three terms in the last summation also represent spin-density coupling and also arise from superexchange. However, unlike the typical superexchange $J$, these terms arise from a second tunneling event which does not reverse the first, but instead allows an atom to hop through its neighbor to an empty site and also flips the spin of the intermediate atom. This ``spin-flip-assisted tunneling'' \cite{jepsen_prx_2021} term affects 3D samples more than 1D because there are 9 times as many next-nearest neighbors.

Finally, we note that even in the absence of explicit spin-density coupling terms, the holes still implicitly couple to spin dynamics by ``reshuffling'' the spin configuration in the sample. This implicit coupling can still destroy the continuous symmetry breaking order required to observe scalable squeezing in 3D.

\section{Polarization Contrast Imaging}

Our system has large atom number, which is associated with already small quantum projection noise relative to the total signal. The signal we intend to measure is the spin imbalance $N_\uparrow - N_\downarrow$, and the quantum projection noise is on the order of $\approx \sqrt{1/10^4} = 1 \%$. If we image two spins separately, as we are subtracting two nearly identical number, any noise and systematic error, such as the $\approx 10 \%$  drift of total atom number, will lead to large error in the measurement. Hence, we opted for imaging the difference in two spin states directly with polarization contrast imaging. We operate at 1028.6G, where the electronic and nuclear spins are nearly decoupled. Thus, we make the approximation that $m_I,m_J$ are good quantum numbers. In reality the electronic and nuclear spins are coupled at the $1-2\%$ level in probability.

\begin{figure}[h]
    \centering
    \includegraphics[width=0.75\linewidth]{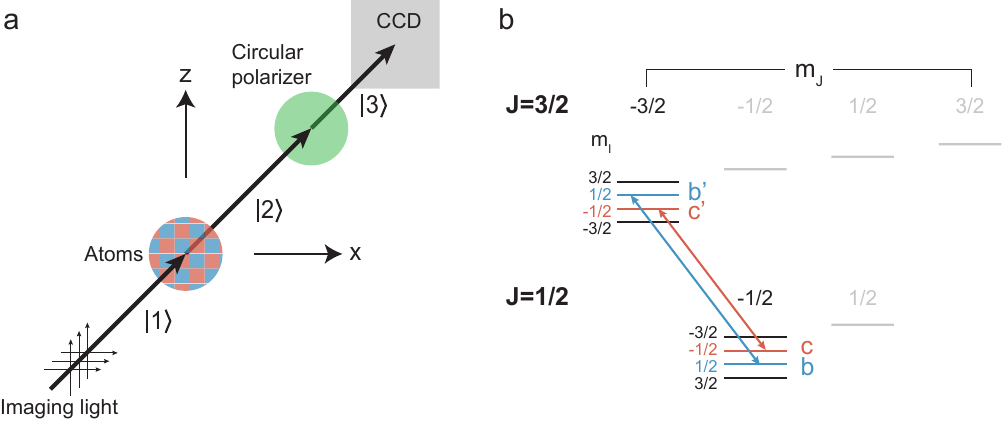}
    \caption{Polarization contrast imaging. (a) Imaging setup. The incoming light is linearly polarized $\ket{1}=(\ket{H}+\ket{V})/\sqrt{2}$. The atoms phase-shift the horizontally polarized light $\ket{2}=(e^{i\vartheta}\ket{H}+\ket{V})/\sqrt{2}$ where the phase $\vartheta$ is proportional to the spin imbalance $S^z=\frac{1}{2}(N_\uparrow-N_\downarrow)$. The light is then projected into the circular basis with a right-handed circular polarizer selecting $\ket{R}$. The final intensity observed at the CCD is $|\braket{R|3}|^2=(1-\sin\vartheta)/2$. (b) Level structure of $^7$Li. Shown are the $m_I$ and $m_J$ components along the quantization axis, which are nearly good quantum numbers at the high magnetic field (1028.55 G) where we operate. The two level system is comprised of the $b$ (blue) and $c$ (red) hyperfine states, which overlap almost completely with the decoupled states $\ket{m_J=-1/2,m_I=\pm 1/2}$. Double headed arrows indicate the allowed transitions for these states with $\sigma^-$ light.}
    \label{fig:imaging}
\end{figure}

Imaging is performed perpendicular to the quantization axis with light in an equal superposition of horizontal and vertical polarizations (Fig.~\ref{fig:imaging}). The horizontally polarized light can be decomposed into $\sigma^+$ and $\sigma^-$ components, while vertically polarized light is $\pi$ polarized. We use the closed $\sigma^-$ transition from $\ket{J,m_J}=\ket{1/2,-1/2}$ to $\ket{3/2,-3/2}$, and park the imaging frequency in between the transition frequency for our two states. Fig.~\ref{fig:imaging} illustrates the imaging setup and level structure of lithium.  We use the $b$ and $c$ states in our experiment. 

\begin{figure}[h]
    \centering
    \includegraphics[width=0.8\textwidth]{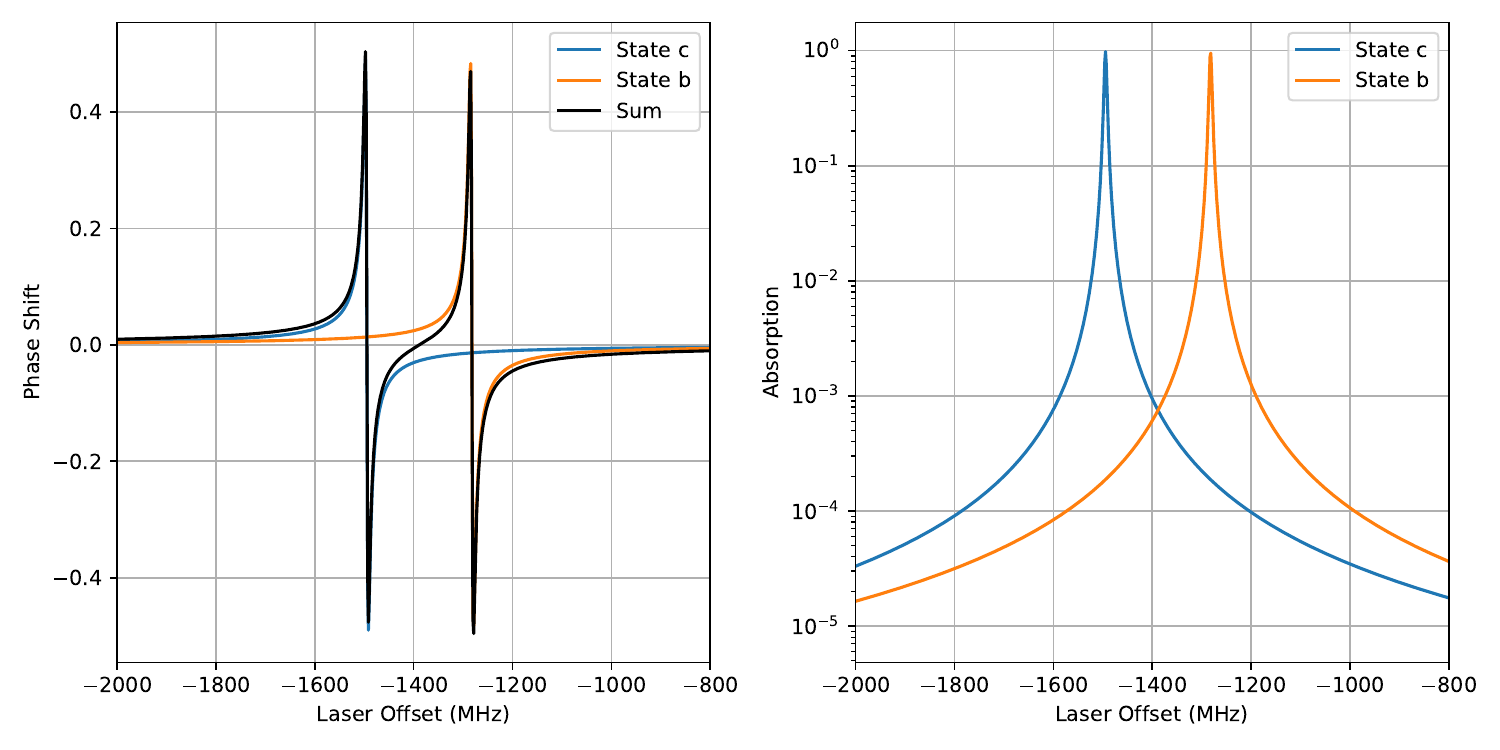}
    \caption{Phase shift imprinted by each atom and residual absorption. At a certain frequency, the phase shifts of atoms in the $\ket{b}$ and $\ket{c}$ states are equal and opposite (left) with minimal absorption (right).}
    \label{fig:imaging_phase_shift}
\end{figure}
What follows is a basic theory of polarization contrast imaging. The imaging frequency for b to b' (c to c') is 1282 (1495) MHz detuned from the D2 resonance centroid at zero field (Fig.~\ref{fig:imaging}). The outgoing electric field ($\ket{H}$ component of $\ket{2}$ of Fig.~\ref{fig:imaging}a) accumulates a phase $\vartheta$ \cite{varenna_notes_bec}
\begin{align}
    \vartheta=-\frac{\tilde{n}\sigma_0}{2}\frac{\delta}{1+\delta^2}
\end{align}
where $\tilde{n}$ is the column density for atoms with detuning $\delta=2(\omega-\omega_0)/\Gamma$ from resonance $\omega_0$ normalized by the half-linewidth $\Gamma/2$, and $\sigma_0=\frac{3}{2\pi}\lambda^2$ is the resonant cross-section for a transition with wavelength $\lambda$. Thus, when the detunings of the imaging light for the b to b' and c to c' transitions are equal and opposite, the phase shift is
\begin{align}
    \vartheta=\frac{\tilde{n}_b-\tilde{n}_c}{2}\frac{\sigma_0}{2}\frac{\delta}{1+\delta^2}
\end{align}
which is proportional to the magnetization in each column $\tilde{n}_b-\tilde{n}_c=\tilde{n}_\uparrow-\tilde{n}_\downarrow$. An extra factor of 2 is included because only the $\sigma^-$ component of the $\ket{H}$ light gains a phase shift, as the $\sigma^+$ component is several GHz detuned from resonance due to the large Zeeman splitting. 

So far we neglected the fact that due to hyperfine interaction, the $b$ and $c$ states we use are not perfect eigenstates of electronic spin. We account this by solving the full hyperfine structure of the ground and excited state of the D2 transition, and the phase shift and absorption per atom is shown in Fig.~\ref{fig:imaging_phase_shift}. 
The point of balanced phase shift is approximately 1388 MHz below the D2 resonance centroid at zero field.
As $\delta\approx35$, the phase shift contributed by each atom is small due to the large detuning. To ensure we are not dominated the photon shot noise, we send approximately $4000$ photons to each atom, resulting in $\lesssim 3$ photons scattered. At the field we operate, the transition is $>98 \%$ closed; with imaging time $<10\mu s$, there is not enough time for heated atom to hop to other sites. We also experimentally verified the apparent loss of atom to dark states or out of trap is less than $10\%$. Hence, the damage during the imaging process is not significant. 

\section{Technical Noise Reduction Techniques}
Our imaging technique based on polarization contrast accounted for atom number fluctuation and the inherent photon shot noise associated with imaging. However, there are various sources of technical or environmental noise. We now explain how we reduced the sensitivity of our measurements to these effects.

\subsection{Fringe removal}
Typically, polarization contrast imaging requires dividing a picture containing the atomic signal by a picture with a reference without atoms (the ``background''). These two pictures are often taken sequentially, and shortening the time between images can mitigate the effects of moving fringes caused by mechanical or acoustic vibrations of optical elements. The speed of the CCD camera limited the time between two images to $<$ 1 ms.  Even for this delay time, moving fringes were detrimental due to the smallness of the signal.  However, thanks to the large number of photons we use, we can remove fringes efficiently using principle component analysis (PCA). 

\begin{figure}[h]
    \centering
    \includegraphics[width=0.8\textwidth]{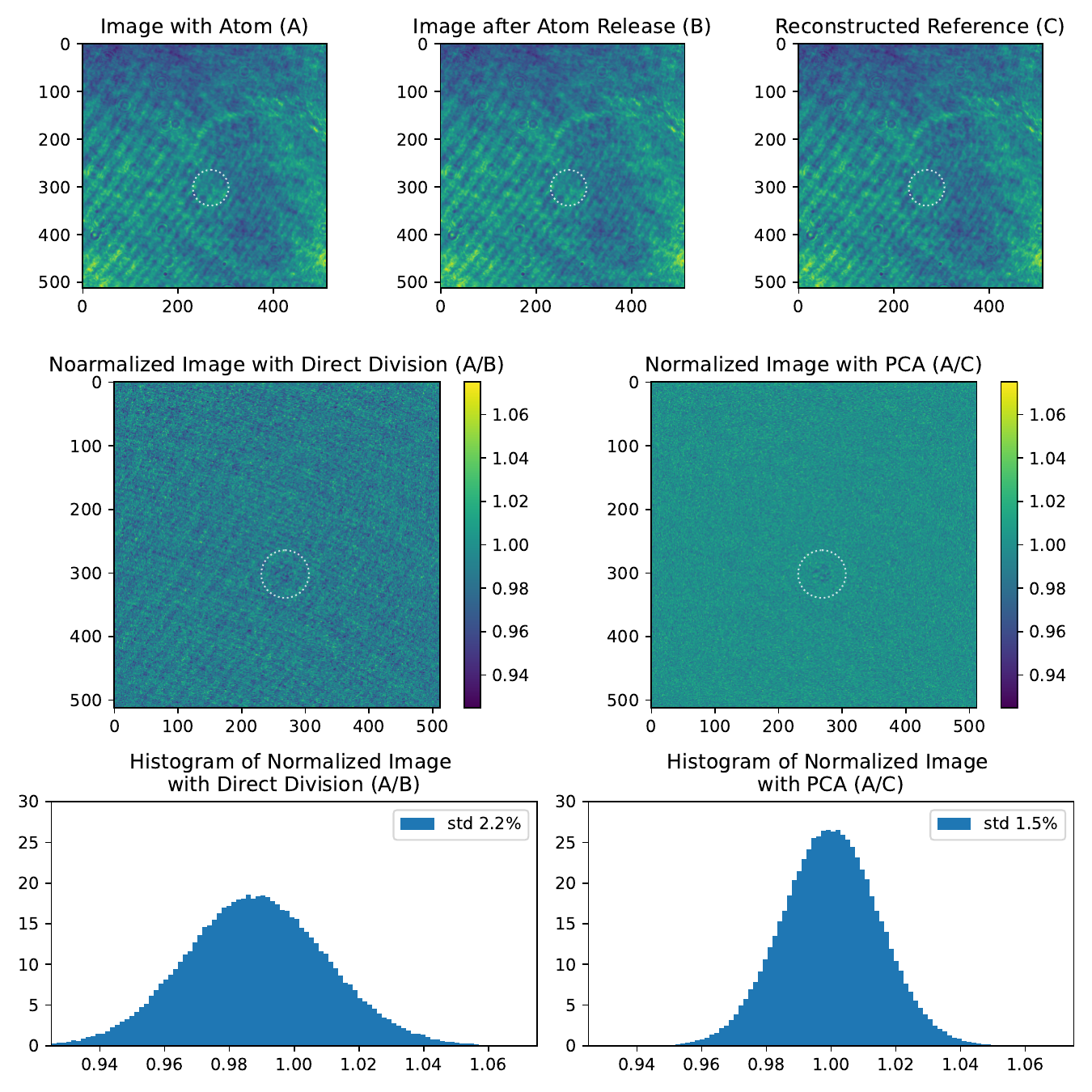}
    \caption{Example of a normalized image produced by principle component analysis (PCA). White dotted lines encloses the location of atoms. The PCA algorithm corrects for imaging intensity fluctuations and fringe movement, which allows the weak signal of spin fluctuations in the sample to be imaged clearly. The residual rms noise of $1.5\%$ (pixels with atoms excluded) is consistent with shot noise arising from the $\sim 5000$ photons each pixel receives.}
    \label{fig:pca_example}
\end{figure}

For each experimental run (which consists of a few thousand cycles), we build a 300-component basis set of drifting fringes using images with no atomic signal. We use this basis of fringes to fit images containing atoms and reconstruct the best possible``background''. This recovers the phase shift imprinted by the atoms with negligible fringes. As shown in Fig.\ref{fig:pca_example}, the residual noise of the reconstructed image is significantly lower, and consistent with the limit arising from photon shot noise.

\subsection{Spin-echo}
Global magnetic field fluctuations affect our Ramsey technique since they introduce shot-to-shot phase fluctuations between the atoms and our local oscillator. Such fluctuation would cause the phase of the analysis pulse to be misaligned from the direction of spin in our sequence. By inserting an echo pulse in the experimental sequence, slowly-varying parts of the noise are removed, while keeping the squeezing intact. 

\subsection{Subsystem analysis}
Although the spin echo can effectively remove slowly-varying magnetic field fluctuations, we found the measurement of ${\rm Var}\,[S^\theta]$ of the entire spin sample in the squeezed state was $\sim$5 SQL. 
Such variances can be attributed to time-dependent magnetic field fluctuations faster than the echo sequence and originates from a variety of sources, including ambient magnetic field, coil thermal expansion, noise in the coil power supply,  the fluctuating vector stark shift caused by power fluctuations of the optical lattice, and small vibrations of the coil induced by cooling water moving at $\sim10$ m/s.
All of these effects give rise to an extra random phase of the spin vector in the equatorial plane between the first and last rf pulses (Fig.~\ref{fig:experiment}b) and therefore extra variances on the order of $N^2\Delta \phi^2$ where $\Delta\phi^2\sim0.01$ rad$^2$ is the typical rms phase accumulation on the scale of our experiments.
Fortunately, all of these noise sources are global on the scale of the atomic sample, and can be removed by analyzing the variance of the differences between two subsystems. In particular,

\begin{align}
    &\mathrm{Var}\left[S^\theta\right]
    =\mathrm{Var}\left[S_a^\theta+S_b^\theta\right]
    =\mathrm{Var}\left[S_a^\theta\right]+\mathrm{Var}\left[S_b^\theta\right]+2\,\mathrm{Cov}\left[S_a^\theta,S_b^\theta\right],\\
    &\mathrm{Var}\left[S_a^\theta-S_b^\theta\right]
    =\mathrm{Var}\left[S_a^\theta\right]+\mathrm{Var}\left[S_b^\theta\right]-2\,\mathrm{Cov}\left[S_a^\theta,S_b^\theta\right],\\
    &\mathrm{Var}\left[S_a^\theta+S_b^\theta\right]\approx\mathrm{Var}\left[S_a^\theta-S_b^\theta\right]. \label{eq:halves_equality}
\end{align}
The last equation holds if the correlation between the two halves is small. This is generally a valid assumption as the maximum entanglement propagation speed is about 1 site per $\hbar/J$ and we operate at times less than $4\hbar/J$. 
All experimental estimates of the variance are calculated using this subtraction method. We note that in the case of squeezing, this method overestimates the global variance; hence, our variance measurements represent a conservative estimate of spin squeezing achieved in the experiment.

\subsection{Four quadrant subsystems}
\begin{figure}[ht]
\centering
\includegraphics[scale=0.7]{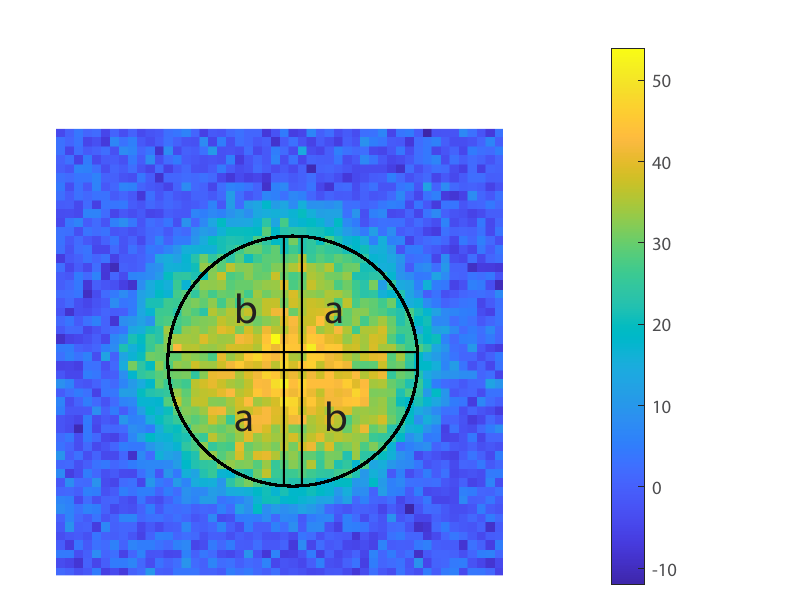}
\caption{Subsystems $a$ and $b$. Region of interest consists of 4 quadrants.  This division  minimizes the extra variance due to the cloud position fluctuations. The gap ensures that the subsystems are independent even for the finite imaging resolution (see below).}

\label{fig:region_of_interest}
\end{figure}

The method of subsystem analysis we have just described removes global magnetic field noise only if the atom number in the two subsystems are equal. Realistically, the position of the atomic cloud fluctuates by 1-2 pixels on the camera from shot to shot. The error in the determination of the center of the cloud leads to an increase of the measured variance. To mitigate this, we divide the cloud into 4 quadrants and form 2 subsystems using the diagonal pairs, as shown in Fig.~\ref{fig:region_of_interest}. The atom number in each subsystem is to first order insensitive to small fluctuation of the sample position. 

\subsection{Double exposure imaging}
We can also correct the position shift of the sample by tracking it in-situ. Although the spin degree of freedom collapses after the spin imbalance measurement, the atom cloud position remains intact. Consequently, we can send another detuned light pulse closer to one of the states to image the position of the atom, as shown in Fig. \ref{fig:experiment}c of the main text. Experimentally we found either the four-quadrant or the double-exposure approach to b sufficient, but used both together in the final analysis.

\subsection{Systematics associated with the subsystem analysis}
\begin{figure}[ht]
\centering
\includegraphics[scale=0.7]{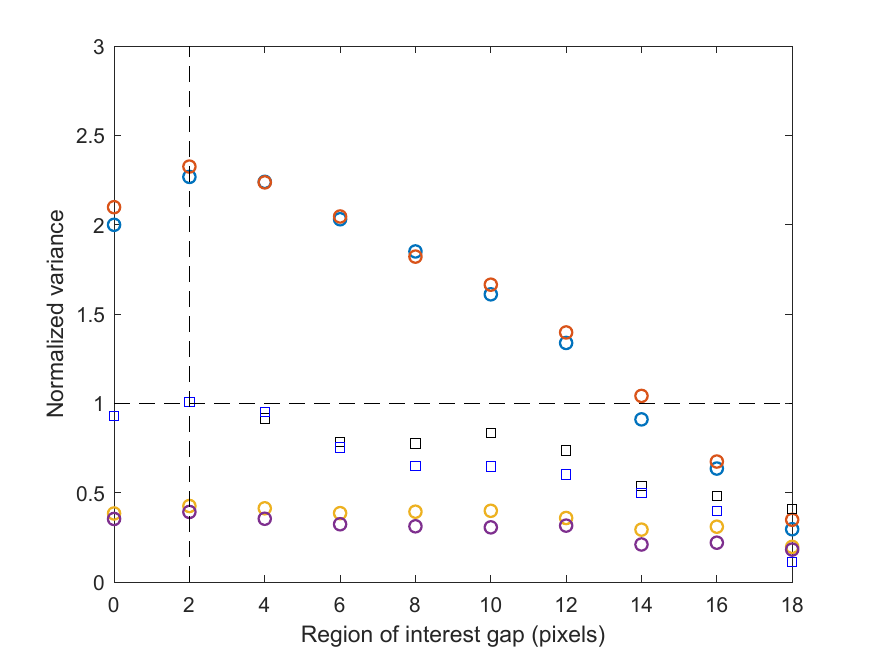}
\caption{Normalized variance vs gap size between the quadrants. The radius of quadrants is fixed to 14 pixels. The gap of 2 pixels (vertical dashed line) is chosen to minimize the bias of variance estimation due to finite optical resolution. Blue (red) circles - anti-squeezing for $\theta=158^o$ ($\theta=153^o$), yellow (purple) - squeezing $\theta=26^o$ ($\theta=31^o$), squares - non-interacting system, all experimental points are for the same evolution time $t=0.7\hbar/J$, horizontal dashed line is expected normalized variance for non-interacting system. }
\label{fig:variance_vs_gap}
\end{figure}

The quadrants are separated by a gap. If the cloud is divided into quadrants with zero gap, finite optical resolution of our detection system (which is approximately 3 $\mu$m or 5 pixels) would decrease the observed variance $\mathrm{Var}[S_a^\theta-S_b^\theta]$, see Fig.~\ref{fig:variance_vs_gap}, because some atoms would be observed in both quadrants 1 and 2.

Therefore the pixels in the gap are excluded from the analysis. Finite optical resolution also leads to a decrease in the measured variance when the size of each quadrant (14 pixels - gap / 2) approaches the optical resolution. This behavior is the same for a squeezed, anti-squeezed and non-interacting cases. Therefore we chose the gap size maximizing the measured variance (2 pixels).

\begin{figure}[ht]
\centering
\includegraphics[scale=0.7]{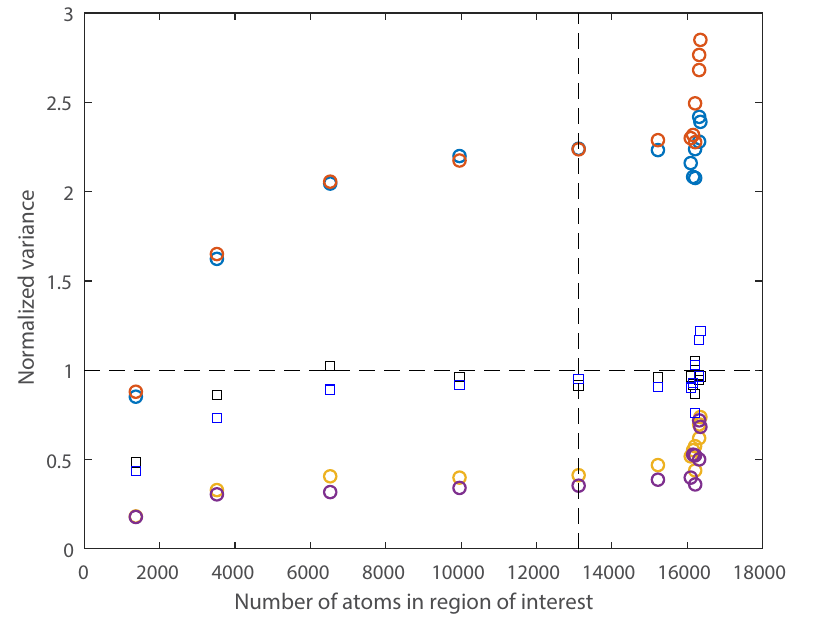}
\caption{Normalized variance vs radius of quadrants. Horizontal axis shows the number of atoms for a given quadrant radius, each point corresponds to radius from 6 pixels to 30 pixels with step of 2 pixels. Radius of the atom cloud is about 17 pixels, vertical dashed line corresponds to radius of 14 pixels chosen for the analysis of all images. Blue (red) circles - anti-squeezing for $\theta=158^o$ ($\theta=153^o$), yellow (purple) - squeezing $\theta=26^o$ ($\theta=31^o$), squares - non-interacting system, all experimental points are for the same evolution time $t=0.7\hbar/J$, horizontal dashed line is expected normalized variance for non-interacting system.}
\label{fig:variance_vs_radius}
\end{figure}

To choose the radius of quadrants we investigated the dependence of the measured variance on the radius of the region of interest (ROI) with a fixed gap of 2 pixels, see Fig~\ref{fig:variance_vs_radius}. Here we plot the variance vs atom number (in the ROI) versus ROI radius from 6 to 30 pixels in steps of 2. The variance becomes noisy when the radius of the ROI approaches the radius of the cloud ($\sim17$ pixels) because the signal per pixel (proportional to atom column density) decreases and some sources of noise, e.g. photon shot noise, are constant. The radius cannot be too small to minimize the effects of variance decrease due to finite optical resolution. Therefore we chose the radius of the quadrants to be 14 pixels corresponding to the vertical dashed line in Fig~\ref{fig:variance_vs_radius}.

\subsection{From image to variance}
We usually average over 200 to 1000 cycles of the experiment with the same parameters for evolution time $t$ and rotation angle $\theta$.
The difference in the camera signal between the two sub-samples were calculated for each shot using the aforementioned methods. We have carefully calculated the expected phase shift per atom (see Fig. \ref{fig:imaging_phase_shift}) and calibrated the magnification of the imaging system by winding a spin helix through angles $0-2\pi$ \cite{jepsen_prx_2021,jepsen_phantom_2022}. 
This allows us to properly normalize the measurement in terms number of atoms and calculate $S_a^\theta-S_b^\theta$ for a given shot. 
Next, the variance of $S_a^\theta-S_b^\theta$ is calculated for the ensemble of shots with identical parameters. 
Besides the noise introduced by atoms, there are also contributions from photon shot noise in the imaging process contributing $\sim0.5$ SQL. 
We correct for this by measuring the variance for shots without atoms, where we apply the same procedure for calculating variances with the atomic signal. 
The normalized variance is calculated as 
\begin{equation}
    4 {\rm Var} [S^\theta] / N\equiv(4\mathrm{Var_{atoms}}[S_a^\theta-S_b^\theta]-\mathrm{4Var_{no\:atoms}}[S_a^\theta-S_b^\theta])/N_{\mathrm{atoms}},
\end{equation}

where $N_{\mathrm{atoms}}$ is the atom number in all 4 quadrants calculated from shots with all spins up in a deep lattice, e.g. without interactions between the atoms.

\section{Simulations of Spin-Density Interactions}

\emph{One dimension--} For the 1D numerics, we employed matrix product state (MPS) methods using the time-dependent variational principle (TDVP) algorithm implemented in ITensor \cite{ITensor-r0.3}. 
We simulated evolution under the full Hamiltonian given by equation~\eqref{eq:tj_full} for $N = 32$ spins and a bond dimension of $\chi = [10,20,40]$.
The Hamiltonian parameters, $\tilde{t}, U_{\sigma\sigma'}$ were determined using the experimental scattering lengths (i.e. were not used as fitting parameters).

We investigated the system evolution with hole densities of $\rho_h=0,0.05,0.11$ using the method described in \cite{jepsen_prx_2021} with $25$ samples.
Briefly, for each sample we initialize the state as $\prod_i (e^{i \theta}\sqrt{p}\ket{h}_i + \sqrt{1-p}\ket{x}_i)$, where $\ket{h}$ and $\ket{x}$ are the (single-site) hole state and $X$ eigenstate, respectively. 
The phase $\theta$ is chosen randomly for each sample, so averaging the results recovers the evolution of the mixed initial state.
We find that for hole fractions of $0\%$ and $5\%$, the spin length and variance observables converge acceptably in bond dimension, with a $5\%$ hole fraction giving better agreement with the experimental data.
The largest hole fraction, $11\%$, does not converge as well, which can be understood as a consequence of larger Hilbert space dimension, $\propto \binom{N}{\rho_hN}$. 
Therefore, it is difficult to assess the density of holes in the experiment solely from the 1D simulations, and densities $5\% \sim 10\%$ have only slight quantitative differences.
Indeed, as discussed in the main text, because the 1D system has no long-range order (even with $\rho_h=0$), the hole fraction does not have a significant impact on the spin length decay or degree of squeezing.
An interesting minor effect is that the spin length \emph{increases} slightly with hole density; this could result from the interaction strength $J$ being effectively diminished due to empty sites.

\begin{figure}[ht]
\centering
\includegraphics[width=0.9\textwidth]{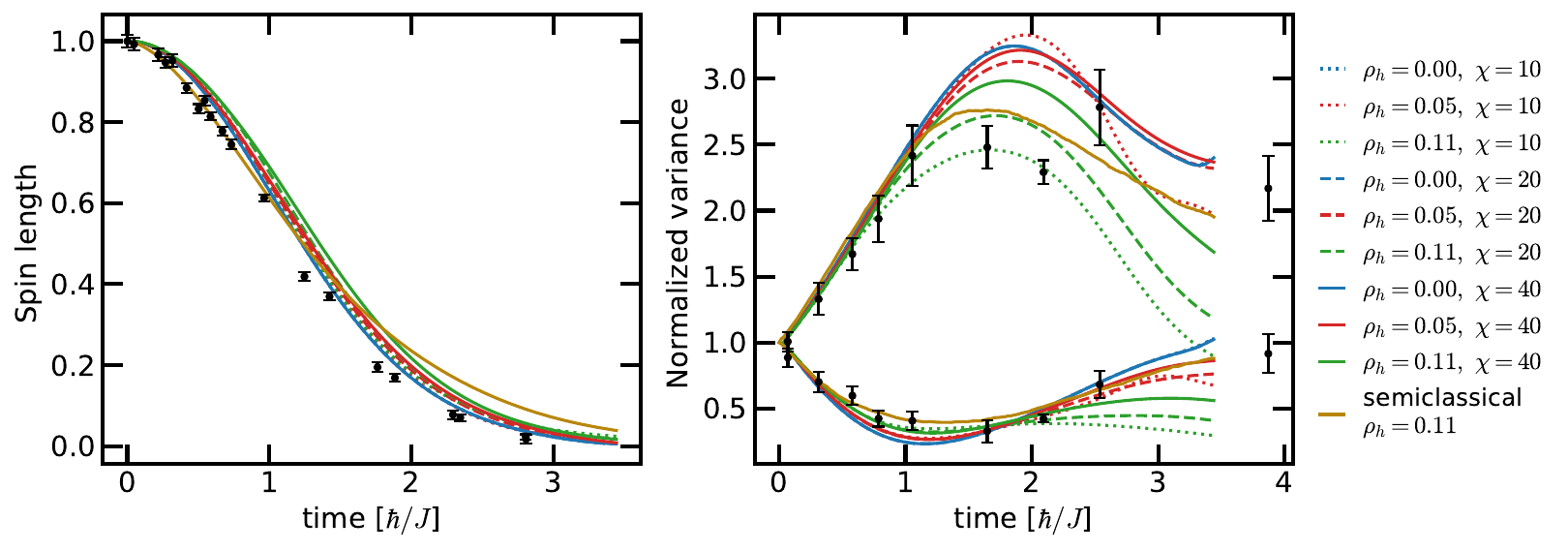}
\caption{Numerical simulations of the 1D experiment using TDVP with varying bond dimension and DTWA with mobile holes (see explanation below). For the TDVP simulations, $dt=0.344$ with $100$ time steps, $N=32$, with $25$ samples. For DTWA, $N=10648$, $dt=0.0172$ with $200$ time steps, and $3000$ samples. Aside from hole density, all simulation parameters were fixed by measured scattering lengths. Overall, the evolution of global observables (contributing to squeezing) depends very little on hole density in 1D. The black symbols are the experimental data.}
\label{fig:t-supp-1d}
\end{figure}

\begin{figure}[ht]
\centering
\includegraphics[scale=0.6]{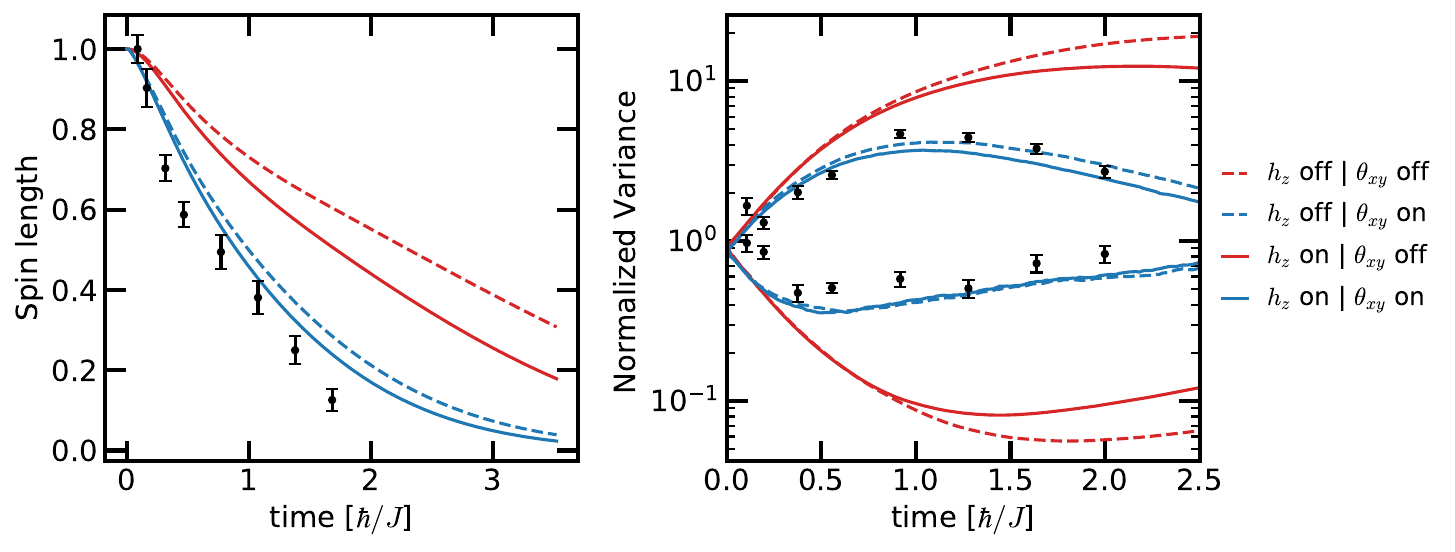}
\caption{Numerical simulations of the 3D experiment using DTWA with mobile holes, with $L=22$, $dt=0.0176$ with $200$ time steps, and $3000$ samples. The simulations were performed with the density-dependent magnetic field and spin-flip assisted coupling terms turned on/off. The spin-flip assisted tunneling term plays a dominant role in causing rapid spin length decay and limiting attainable spin-squeezing. Aside from hole density (see discussion below), all simulation parameters were fixed by measured scattering lengths.}
\label{fig:t-supp-3d}
\end{figure}

\emph{Three dimensions--} For three dimensions, exact quantum calculations are no longer possible. Therefore, we adopted a semi-classical approach.  The discrete truncated Wigner approximation (DTWA) allows for the efficient simulation of large spin systems while capturing the quantum correlations leading to spin squeezing \cite{dtwa_rey_2015,theory_shortrange_2020,maxbingtian}. 
To incorporate the effect of mobile holes, we augmented DTWA with additional dynamics beyond the standard Hamiltonian evolution. 
In detail, the simulation is performed as follows:
\begin{enumerate}
    \item Initialize the classical analogue of $H_0$. Rescale such that $J=1$.
    \item Initialize the system with classical spins drawn from $\vec{S}_i = {(1/2, \pm 1/2, \pm 1/2)}$. 
    \item Assign $\rho_h N$ random sites to be holes, by replacing $\vec{S}_i \to \vec{0}$
    \item Repeat steps 5-10 for $N$ time steps
    \item Add single-site $z$-directed magnetic fields of strength $h_z$ to any site that is the nearest neighbor of a hole.
    \item Evolve the spin ensemble under the classical equations of motion for $dt$
    \item Remove the $h_z$ from step 5
    \item For each hole, exchange it with a nearest neighbor spin with probability $\tilde{t}dt$.
    \item For each hole, exchange it with a nearest neighbor spin \emph{and then} a next-nearest neighbor spin with probability $(2z-1)dt$.
    \item For the intermediate spin from step 9, $\vec{S}_j$, apply the following rotation: sample a random direction in the $xy$ plane $\hat{n}_{xy}$ and a random angle $\theta_{xy} \in [0,2\pi]$ and set $\vec{S}_j \to R(\hat{n}_{xy},\theta_{xy})\vec{S}_j$.
    \item Save total spin expectation values.
\end{enumerate}
Steps 3,5,6 are straightforward classical approximations of the $\sim S^z_i n_{j\sigma}$ terms of the full Hamiltonian given by equation~\eqref{eq:tj_full}. 
Step 8 accounts for the direct hopping term, and is normalized such that holes undergo $\tilde{t}$ nearest-neighbor exchanges in unit (i.e. $\sim 1/J$) time.
Steps 9 and 10, which emulate the spin-flip assisted tunneling and are crucial for obtaining agreement with experiment (see Fig.~\ref{fig:t-supp-3d} and discussion below), merit further explanation.
The essential idea is that we approximate the density degree of freedom as a classical random variable that has no persistent entanglement with the spins, reminiscent of the Born-Markov approximation. 
Under this assumption, we have ${\rm Tr}_{ik}[a^{\dagger}_{i\bar{\sigma}}S^{\sigma}_j a_{k\sigma}] \approx c^{\sigma\bar{\sigma}}_{ik}(t) S^{\sigma}_j$ where $c^{\sigma\bar{\sigma}}_{ik}(t)=\langle a^{\dagger}_{i\bar{\sigma}} a_{k\sigma} \rangle$ is the hopping correlation function.
Suppressing the indices on the correlator, the total spin-flip assisted tunneling term is $c S^{+}_j + c^* S^{-}_j = 2{\rm Re}(c) \; S^x_j + 2{\rm Im}(c) \; S^y_j$.
Therefore, in a semi-classical approximation, the spin-flip assisted tunneling should effect a rotation on site $j$ about an $xy$ axis determined by the phase of $c$.
In our approximation, the holes do not have coherence over long time scales, i.e. $\frac{1}{T} \int_0^T c(t) dt \approx 0$, so $c(t)$ must have a time varying phase implying a random axis of rotation.

It is more difficult to estimate the appropriate angle, but the coefficient of this term is $\frac{J}{4}$ so an $O(\pi)$ rotation should occur every time step.
However, the strength of the rotation should fluctuate, since the hole-correlation function itself fluctuates.
According to our stipulation that the density and spin should never remain entangled, the rotation has to occur instantaneously, i.e. during a $dt$ step.
Taken together, these considerations indicate the rotation should be applied upon a ``double hopping" event and the rotation angle should be drawn from $[0,\alpha 2\pi]$ where $\alpha$  is not too small.
In practice, we found the value of $\alpha$ was unimportant and fixed $\alpha=1$.
A final point is that the ``double hop" event occurs once per unit time, with an additional factor of $2z-1$ to account for the multiplicity of hopping pathways in higher-coordination lattices (the 2nd hop can be to any of $2z$ neighbors except back to the original site).
This coordination factor contributes to the greater significance of holes in 3D than 1D.
In fact, as shown in Fig.~\ref{fig:t-supp-1d}, applying the above semiclassical simulation to the 1D system with $\rho_h=0.11$, we obtain similar results to the TDVP numerics, and the large hole fraction has little effect on the dynamics.

\begin{figure}[ht]
\centering
\includegraphics[scale=0.6]{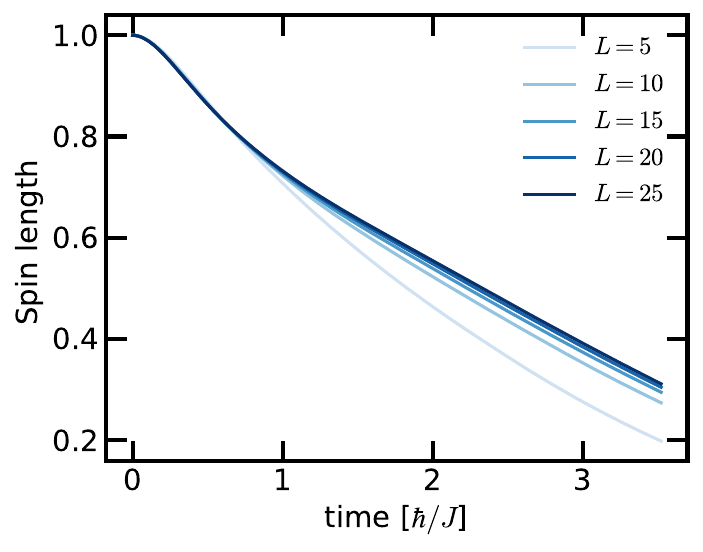}
\caption{Simulations of varying lattice sizes $L=[5\cdots 25]$ in 3D using DTWA with holes to investigate the stability of CSB with $\rho_h=0.11$. For all system sizes, $dt=0.0176$, with $200$ time steps, and $2000$ samples. Here, the density-dependent magnetic field and spin-flip assisted coupling terms are turned \emph{off}. The spin length decay clearly converges to a rapid, constant decay with system size, indicating the absence of long-range order at this hole density.}
\label{fig:t-supp-shuffle}
\end{figure}

The situation in 3D is very different.
To begin with, even \emph{without} the direct spin-density coupling (i.e. setting $h_z=0$ and $\theta_{xy}=0$ in the numerics), the presence of holes still leads to system-size independent spin length decay, meaning the system does \emph{not} exhibit continuous symmetry breaking (Fig.~\ref{fig:t-supp-shuffle}).
This shows that the initial hole density is sufficient to disorder the system, but it does not quantitatively explain the rapid spin length decay and limited squeezing of the experiment.

To understand which terms of $H$ [eq.~\eqref{eq:tj_full}] are most adverse for squeezing, we turn on the effects of $h_z$ and $\theta_{xy}$ independently.
The results are shown in Fig.~\ref{fig:t-supp-3d}; the most striking feature is the strong effect of $\theta_{xy}$ -- independent of $h_z$, it is almost entirely responsible for giving quantitative agreement with the experiment.
The somewhat unexpected significance of this second-order term highlights the need to develop a complete and nuanced understanding of the role of density defects in ultracold atomic lattice systems.
We note that although the DTWA simulations clearly show the importance of spin-density coupling, they do not necessarily provide an accurate estimate of the hole density to the few $\%$ level; indeed, the semiclassical approximation inherently neglects terms of the form $\sim a^\dag_k n_j a_k$ in equation~\eqref{eq:tj_full}, and hence may \emph{overestimate} the density of holes needed to recover the observed spin length decay.
Therefore, the actual hole density of the sample is likely closer to the $5\%$ estimate given by the more rigorous TDVP simulations.

A crucial challenge for future research is to determine the critical hole density $\rho_h^*$ such that $\rho_h < \rho_h^*$ features continuous symmetry-breaking (CSB) order and enables scalable squeezing.
We emphasize this is not possible with the semiclassical numerics presented here: this algorithm is only accurate at short-times, when the density degrees of freedom are much hotter than the spin degrees of freedom and they have not yet reached thermal equilibrium.
While this limitation is not an issue for analyzing the present experiment, which is confined to relatively short evolution time, future efforts to achieve scalable squeezing will require numerics that are accurate to at least $O(N^{1/3})$ time.
This requires performing Monte Carlo sampling on the possible hole jumps \emph{during} the dynamics to maintain detailed-balance.

\newcounter{sbib}
\setcounter{sbib}{35}


\begin{thebibliography}{10}
\bibitem{footnote_system_sizes} We note that at sufficiently large system sizes, the implicit coupling does arrest scalable squeezing \cite{supp}.
\end{thebibliography}
\end{document}